# Influence by small dispersive coal dust particles of different fractional consistence on characteristics of iodine air filter at nuclear power plant

I. M. Neklyudov, O. P. Ledenyov, L. I. Fedorova, P. Ya. Poltinin

*National Scientific Centre Kharkov Institute of Physics and Technology,
Academicheskaya 1, Kharkov 61108, Ukraine.*

The main purpose of research is to determine the influence by the small dispersive coal dust particles of the different fractional consistence on the technical characteristics of the vertical iodine air filter at nuclear power plant. The research on the transport properties of the small dispersive coal dust particles in the granular filtering medium of absorber in the vertical iodine air filter is completed in the case, when the modeled aerodynamic conditions are similar to the real aerodynamic conditions. It is shown that the appearance of the different fractional consistence of small dispersive coal dust particles with the decreasing dimensions down to the micro- and nano- sizes at the action of the air-dust aerosol stream normally results in a significant change of distribution of the small dispersive coal dust particles masses in the granular filtering medium of an absorber in the vertical iodine air filter, changing the vertical iodine air filter's aerodynamic characteristics. The precise characterization of the aerodynamic resistance of a model of the vertical iodine air filter is completed. The comparative analysis of the technical characteristics of the vertical and horizontal iodine air filters is also made.

## Introduction

In the early experimental research to characterize a model of the absorber with the similar aerodynamics characteristics as in the case of the absorber in the iodine air filter (*IAF*) at the nuclear power plant (*NPP*), we found that there is a sharp increase of the magnitude of the *IAF* aerodynamic resistance (in more than *20* times) during its operation in the regime of air filtering at the *NPP* over several years [1]. We researched the failed *IAF* and determined that the small dispersive coal dust particles with the various geometric shapes and dimensions down to the micro- and nano- sizes were accumulated in the absorber in the *IAF* over many years of its operation, resulting in both the sharp increase of magnitude of absorber's aerodynamic resistance and the subsequent *IAF* failure. At the same time, the interconnection between the appearance of observed phenomena and the fractional consistence of small dispersive coal dust particles, representing the crushed cylindrical coal granules of absorbent of the type of *CKT-3* in the sub-surface layer of absorber, was not well understood. It became clear that the cylindrical coal absorbent granules of the type of *CKT-3* were crushed in the process of the turbulent air-dust aerosol stream action, generating the small dispersive coal dust particles with the decreasing dimensions down to the micro- and nano- sizes, which were captured and moved forward by the air-dust aerosol stream inside the *IAF* at the *NPP*. It is necessary to mention that, during the process of experimental modeling of absorber operation with the application of the artificially synthesized small dispersive coal dust particles of the different sizes in [1], the small dispersive coal dust particles were synthesized in the process of forced crushing of the cylindrical coal adsorbent granules of the type of *CKT-3* inside a special centrifugal mill, operating at the small velocities of rotation. The obtained small dispersive coal dust particles had the fractional consistence, which was similar to the fractional consistence, appearing at the sub-surface layer of absorber in the real *IAF* at the *NPP*. During the detailed research analysis of the obtained experimental data in [2], the series of maximums of coal dust masses was discovered along the absorber in the *IAF*. The maximums were situated in the close proximity to the sub-surface layer of absorbent as well as deeply inside the absorbent. We were able to explain the physical mechanism of origination of the coal dust masses maximums by analyzing the small dispersive coal dust particles origination and movement dynamics in the real absorbers in the *IAFs* at the *NPP*, taking to the consideration: 1) the interaction between the small dispersive coal dust particles and the cylindrical coal adsorbent granules, as well as 2) the interaction of the small dispersive coal dust particles with the other small dispersive coal dust particles in the process of air-dust aerosol stream flow in the *IAF* at the *NPP* [2]. It was found that the small dispersive coal dust particles of large sizes originate the creation of the first coal dust masses maximum in the sub-surface layer of absorber;

here, it makes senses to note that the small dispersive coal dust particles of all the sizes participate in the creation of this maximum. In our opinion, the appearance of the dense monolithic dust layer, made of the small dispersive coal dust particles of all the sizes, is a main reason of the subsequent exponential increase of the magnitude of the *IAF* aerodynamic resistance at the *NPP*. We would like to emphasis that every next coal dust mass maximum inside the absorber was created by the small dispersive coal dust particles of the certain discrete dimension. It is important to explain that the fractions of the small dispersive coal dust particles with the smaller sizes participate in the creation of the clots, which are situated on the bigger distances from the absorber's input surface, comparing to the positions of the clots, which are originated by the small dispersive coal dust particles of the large sizes. At the action by the air-dust aerosol stream flow, the coal dust masses maximums shift along the length of an absorber and can be even jettisoned outside the *IAF*. In [3], it was discovered that there is a significant change in the physical character of the absorption process of both the radioactive chemical elements (*Iodine*, *Barium*, *Strontium*) and the radioactive isotopes, because of the presence of the discovered coal dust masses maximums, namely the selective absorption of the radioactive chemical elements (*Iodine*, *Barium*, *Strontium*) and the radioactive isotopes was observed, depending on the dimensions of the small dispersive coal dust particles.

Going from the observed distinction of character of transportation of both the small dispersive coal dust particles of the large sizes and the small dispersive coal dust particles of the small sizes in [2], we proposed that the change of the directions between the gravitation force direction and the air-dust aerosol stream flow direction can result in both 1) the change of distribution of the small dispersive coal dust particles masses along the length of an absorber inside the *IAF*, and 2) the possible improvement of the aerodynamic characteristics of the *IAF*. The conducted aerodynamic tests with a model of the horizontal *IAF* [4] allowed us to find that its magnitude of aerodynamic resistance increases in the *1,7* times, comparing to the increase of the magnitude of aerodynamic resistance in the *20* times approximately in a model of the vertical *IAF* at the same limiting quantity of the introduced coal dust mass share. The similar effect can be explained by the fact that the small dispersive coal dust particles of the large sizes, which originate the processes of creation and structurization of the small dispersive coal dust particles clots in the sub-surface layer of the granular filtering medium of an absorber, continue to shift to the bottom in the horizontal *IAF* at the action by the gravitation force, as a result the particles cannot create the homogeneous blocking layer in the granular filtering medium of an absorber in a model of the horizontal *IAF*. The obtained research results allow us to suppose that the change in the fractional consistence of the small dispersive coal dust particles from the large sizes to the small sizes (below than *1 µm*), must strongly decrease the influence by the small dispersive coal dust particles fraction on the magnitude of the aerodynamic resistance of the vertical *IAF*. Therefore, this research is focused on the study of both 1) the distribution of the small dispersive coal dust particles with the small sizes below *1 µm* along the length of absorber in the vertical *IAF* and 2) the influence by the small dispersive coal dust particles fractional consistence on the coal dust particles transport properties in the granular filtering medium in the absorber in a model of the vertical *IAF*.

## Experimental measurements technique

The introduced small dispersive coal dust particles were created in the process of crushing of absorbent made of the cylindrical coal granules of the type of *CKT-3* with the application of a special centrifugal mill. The distribution of the small dispersive coal dust particles dimensions was shifted to the range of small sizes, comparing to the small dispersive coal dust particles dimensions, which were used in [1].

In Fig. 1, a model of the vertical *IAF* with the ten containers, divided by the nets with the large cells, is shown. The source-container with the mixture of the small dispersive coal dust particles and the cylindrical coal granules (not more than *1,5 %* of the mass of the small dispersive coal dust particles from the total mass of the absorbent) was situated in front of the ten containers. The source-container was re-charged with the new mixture of the small dispersive coal dust particles and the cylindrical coal granules before the beginning of every next experiment. The ten containers were filed with the homogeneous cylindrical coal granules of the type of *CKT-3* with the length ~ *3,2 mm* and the diameter ~ *1,8 mm*, creating the granular filtering medium in the absorber in the vertical *IAF*.

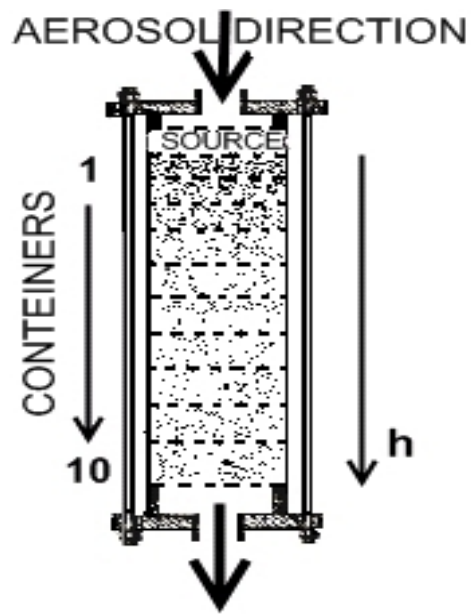

***Fig. 1.*** *Scheme of model of vertical iodine air filter, including source-container with small dispersive coal dust particles of small sizes and ten containers with granular filtering medium with cylindrical coal granules of type of CKT-3.*

The precipitated mass of the small dispersive coal dust particles was determined as a difference between the mass of a container after the completion of experiment and the mass of a container before the beginning of experiment. The measurement of the precipitated mass of the small dispersive coal dust particles was conducted for every of the ten containers.

The following technical designations are used:
*$M_0$* is the mass of absorbent in the absorber;
*$M_i$* is the mass of absorbent in the *i*-section of the absorber (*i* changes from *1* to *10*);
*$m_0$* is the total mass of the small dispersive coal dust particles, which was precipitated in the absorber at the experiment completion;
*$m_i$* is the mass of the small dispersive coal dust particles, which was precipitated in the *i*-section of an absorber at the experiment completion;
*$h$* is the length of a model of the *IAF*.

It is necessary to clarify that the difference between the input pressure and the output pressure in a model of the *IAF* was measured by the manometer.

### 3. Experimental measurements results
### 3.1. Distribution of small dispersive coal dust particles in granular filtering medium along length of absorber in vertical iodine air filter

In Fig. 2, the experimental modeling results on the distribution of the precipitated mass of the small dispersive coal dust particles along the length of an absorber (see the graph 1 in Fig. 2) and the averaged data in [1] (see the graph 2 in Fig. 2) are shown.

As it can be seen from the graphs 1 and 2 in Fig. 2, the mass share of the coal dust fraction in the granular filtering medium in the sub-surface layer of *2 cm* is $m_i/(M_i+m_i)=18,75\%$. As it can be noted in the graph 2 in Fig. 2, the clearly defined maximums of the coal dust masses, which are characteristic for the case of the small dispersive coal dust particles of the large sizes, were not detected in this experiment.

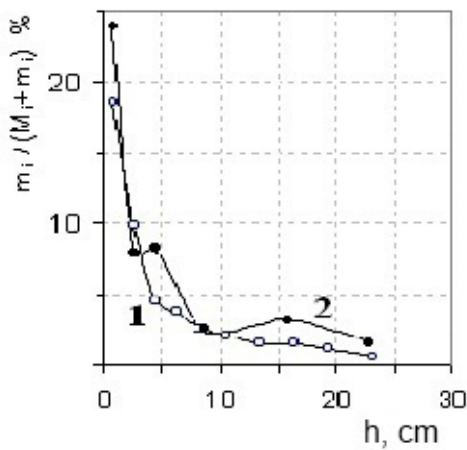

***Fig. 2.*** *Dependence of distribution of mass share of coal dust fraction in granular filtering medium along length of absorber in model of vertical IAF at end of series of experiments ($m_i/(M_i+m_i)$) (h). Dimensions of small dispersive coal dust particles: 1– smaller than 1 µm; 2 – smaller than 10 µm; i − from 1 to 10 µm.*

We found that the coal dust mass share sharply decreases inside an absorber in a model of the vertical *IAF*. Going from our experimental measurements, the coal dust mass share decreased in the four times on the distance of *5 cm* in an absorber in a model of the vertical *IAF*. This dependence can be well approximated by the expression

$$m_i/(M_i+m_i) = 18{,}75 \cdot (1-\mathrm{erf}z),$$

where:
***erfz*** is the *Gauss* "errors integer" in [2];
$z = (3)1/2 \cdot (h-0{,}89286) / 2 (Dt)1/2$;
*$h$* is the distance, which is measured from the input surface of the absorber in a model of the *IAF* in the direction of the air-dust aerosol stream flow (see the graph 2 in Fig. 3);
*$D$* is the effective coefficient of diffusion;
*$t$* is the time period of experiment completion.

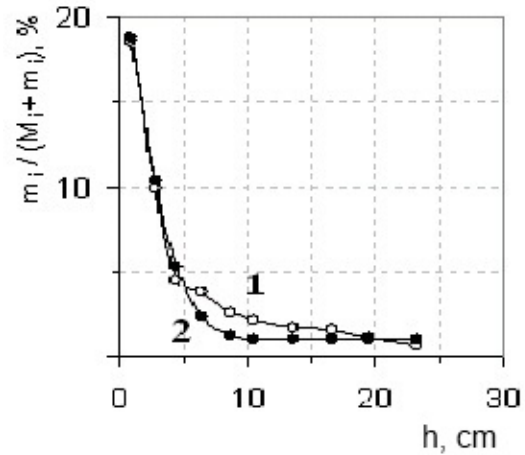

***Fig. 3.*** *Dependence of distribution of mass share of coal dust fraction in granular filtering medium along length of absorber in model of vertical IAF ($m_i/(M_i+m_i)$) (h)*
*1 - experimental dependence;*
*2 – computer modeling dependence.*

In the graph 1 in Fig. 3, the experimental data show the smooth decrease of the coal dust mass share below *0,6%* at the output of the *IAF*. The mass of the small dispersive coal dust particles, which was jettisoned outside the *IAF*, is determined as a difference between the limiting mass of the small dispersive coal dust particles, which was introduced in the absorber, and the total mass of the small dispersive coal dust particles, which was precipitated in the granular filtering medium in the sections in the absorber. In our experiment, the jettisoned mass of the small dispersive coal dust particles was *61 %* of the introduced mass of the small dispersive coal dust particles. In the experiment with the small dispersive coal dust particles of the large sizes [1], the jettisoned mass of the small dispersive coal dust particles was *44 %* of the introduced mass of the small dispersive coal dust particles.

## 3.2. Influence by fractional consistence of small dispersive coal dust particles on magnitude of aerodynamic resistance in vertical iodine air filter

In the experiment, the dependence of the aerodynamic resistance on the volumetric air stream flow $\Delta P(J)$ in the case of the increasing values of the coal dust masses concentration share, $m_o/(M_o+m_o)$, which was introduced in a model of the vertical *IAF*, was researched. In Fig. 4, the graphs of dependence of the aerodynamic resistance on the volumetric air stream flow $\Delta P(J)$ at the different values of the coal dust mass share, $m_o/(M_o+m_o)$, are shown. Going from the comparative analysis of the obtained experimental results, it is possible to make a conclusion that the magnitude of the aerodynamic resistance of a model of the *IAF*, which was enriched by the small dispersive coal dust particles of the large sizes as well as the magnitude of the aerodynamic resistance of a model of the *IAF*, which was enriched by the small dispersive coal dust particles of the small sizes, synchronously increase up to the value of relative coal dust mass share $m_o/(M_o+m_o) = 7\ \%$ (see the graphs 2 and 5 in Fig. 4), then the magnitude of the aerodynamic resistance of a model of the *IAF*, which was enriched by the small dispersive coal dust particles of the small sizes, continues to increase slowly in comparison with the fast increase of the magnitude of the aerodynamic resistance of a model of the *IAF*, which was enriched by the small dispersive coal dust particles of the large sizes (see graphs 8 and 3, 9 and 6 in Fig. 4). In the case of the aerodynamic resistance $\Delta P = 6000\ Pa$, the magnitudes of the aerodynamic resistances $\Delta P^*$, which are normalized to the constant air stream $J^*=15\ m^3/hour$, are calculated, using the early obtained empirical dependence of the aerodynamic resistance on the volumetric air stream $\Delta P(J)$ in [1] and the data in the graphs in Fig. 4.

dust particles: *1 - 9* are below *1μm*, *1 - 6* are below *10 μm*.
○ - *1 (0), 2 (0,2), 3 (1,4), 4 (4,3), 5 (5,8), 6 (6,7),7 (7,4), 8 (8,4), 9 (9,3)*;
● - *1 (0), 2 (5,9), 3 (6,7), 4 (8,8), 5 (9,1), 6 (9,2)*.

In Fig. 5, the dependence of the aerodynamic resistance at the air stream $J^*=15\ m^3/hour$ on the relational mass share of coal dust fraction in the vertical *IAF*, is shown. As it can be seen from the comparison, the magnitude of the aerodynamic resistance of a model of the vertical *IAF*, which was enriched by the small dispersive coal dust particles of the large sizes as well as the magnitude of the aerodynamic resistance of a model of the vertical *IAF*, which was enriched by the small dispersive coal dust particles of the small sizes, synchronously increase up to the value of relative coal dust mass share $m_o/(M_o+m_o) = 7\ \%$ (see the graphs 1 and 2 in Fig. 5). Then, the magnitude of the aerodynamic resistance of a model of the vertical *IAF*, which was enriched by the small dispersive coal dust particles of the large sizes, sharply increases in more than *20* times, comparing to the initial magnitude of the aerodynamic resistance (see the graph 2 in Fig. 5). The magnitude of the aerodynamic resistance of a model of the vertical *IAF*, which was enriched by the small dispersive coal dust particles of the small sizes, continues to increase linearly and it approximately equals to $\Delta P \approx 4500\ Pa$ at the coal dust mass share of *9,3 %*, that is in the *2* times bigger than the initial magnitude of the aerodynamic resistance. Comparing the two cases, it is necessary to emphasis that, at the completion of the experiment, the magnitude of the aerodynamic resistance of a model of the vertical *IAF*, which was enriched by the small dispersive coal dust particles of the large sizes was in the ten times higher than the magnitude of the aerodynamic resistance of a model of the vertical *IAF*, which was enriched by the small dispersive coal dust particles of the small sizes as shown in the graphs 1, 2 in Fig. 5.

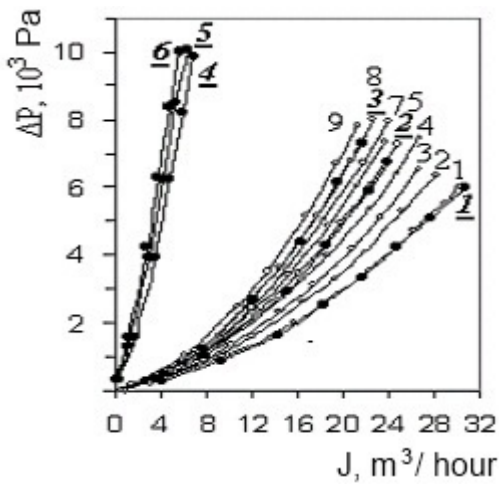

***Fig. 4.*** *Dependence of aerodynamic resistance of absorber as function of volumetric air stream flow $\Delta P(J)$ in cases of different values of mass share of introduced small dispersive coal dust particles mo/(Mo+mo) (%) in vertical IAF. Dimensions of coal*

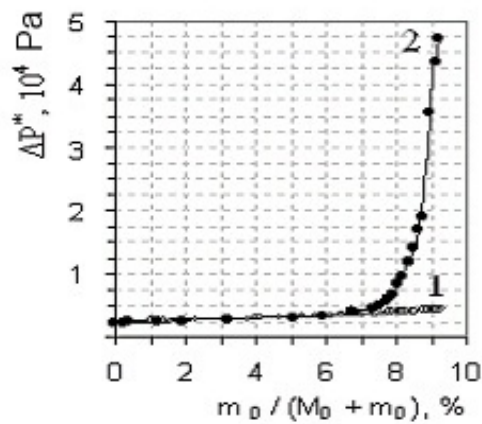

***Fig. 5.*** *Dependence of aerodynamic resistance at air stream $J^*=15\ m^3/hour$ on relational mass share of coal dust fraction, captured in iodine air filter.*
*1 - absorber was enriched by small dispersive coal dust particles of small sizes;*
*2 – absorber was enriched by small dispersive coal dust particles of large sizes.*

Going from the comparative analysis of data in the graph 1 in Fig. 2 and in the graph 1 in Fig. 5, it is possible to conclude that: 1) the clot made of the small dispersive coal dust particles of the small sizes can be penetrated by the air stream in the vertical *IAF*; and 2) the clot made of the small dispersive coal dust particles of the large sizes cannot be penetrated by the air stream in the vertical *IAF*. Therefore, there is no an exponential increase of the aerodynamic resistance and subsequent air-dust aerosol stream flow blocking in the case of the small dispersive coal dust particles of the small sizes [1]. Let us note that the relative quantity of the small dispersive coal dust particles of the small sizes, which was jettisoned by the air-dust aerosol stream flow outside the vertical *IAF*, is in the *1,5* times higher, comparing to the case the small dispersive coal dust particles of the large sizes.

## Conclusion

The research on the transport properties of the small dispersive coal dust particles in the granular filtering medium of absorber in the vertical iodine air filter is completed in the case, when the modeled aerodynamic conditions are similar to the real aerodynamic conditions. It is shown that the origination of the different fractional consistence of the small dispersive coal dust particles with the decreasing dimensions down to the micro- and nano- sizes due to the action by the air-dust aerosol normally results in a significant change of the distribution of the small dispersive coal dust particles masses in the granular filtering medium of an absorber in the vertical *IAF*, changing its aerodynamic characteristics. The experimental results on the precise characterization of the aerodynamic resistance of the vertical *IAF*, in the case of a strong influence by the small dispersive coal dust particles, are obtained. In the experiments with the small dispersive coal dust particles fraction, it was found that a shift in the fractional consistence of the small dispersive coal dust particles toward an increased number of particles with the very small sizes results in an immediate suppression of the exponential increase of aerodynamic resistance magnitude at the application of increased integral dust loads in the vertical *IAF*. It is necessary to remind that these big integral dust loads in the absorbers in the *IAF* were considered as a main reason of the absorber failures in the *IAF* at the *NPP*. In the researched case, the aerodynamic resistance of a model of the vertical *IAF* slowly increases in accordance with the linear law. At the accumulation of the coal dust mass share up to *9,3 %* in the absorbers in the vertical *IAF*, the aerodynamic resistance magnitude only increases up to *4500 Pa* (in the *2* times) in the case of the small dispersive coal dust particles of the small sizes; while the aerodynamic resistance magnitude increased up to *45000 Pa* (in the *20* times) in the case of the small dispersive coal dust particles of the large sizes. The shape of distribution of accumulated coal dust mass share along the length of an absorber has a one maximum, situated in the sub-surface layer of absorber in the vertical *IAF*.

The obtained research results can be theoretically explained, going from the consideration of the physical mechanisms, which have an influence on the transport properties of the small dispersive coal dust particles in the granular filtering medium of an absorber in the vertical *IAF*. In agreement with [2], the small dispersive coal dust particles movement in the absorber in the *IAF* is a complex physical process, which is accompanied by the elastic and non-elastic collisions by the particles with the cylindrical absorbent granules, resulting in the changes of their kinetic energies. The air-dust aerosol stream flow has place in the system of air channels with the variable diameters between the cylindrical coal granules in the granular filtering medium in a model of the vertical *IAF*. The presence of the empty cavities between the cylindrical coal granules, where the velocity of the air-dust stream flow is relatively slow, mainly results in an accumulation of the small dispersive coal dust particles of the large sizes in the sub-surface layer of the absorber in the vertical *IAF*. The small dispersive coal dust particles of the micro- and nano- sizes are usually transported on the long distances along the length of an absorber, creating the subsequent coal dust masses concentration maximums, but they can also precipitate on the large sizes particles, which are anchored in the sub-surface layer of the granular filtering medium in an absorber in the vertical *IAF*. The next coal dust masses concentration maximums are originated due to the fact that there is a maximal exchange by the energies and the impulses in the case, when the small dispersive coal dust particles with the equal masses are collided elastically. The smaller dispersive coal dust particles of the micro- and nano- sizes can partly penetrate through the existing coal dust masses maximums and create their next maximums, which are situated on the certain distances from the previous maximums. All these maximums, except for the first one, have the small drift velocities with the movement vectors direction, which coincides with the direction of the air-dust aerosol stream flow, hence they constantly shift along the length of an absorber in the vertical *IAF* over the time [2]. In view of the fact that we used the small dispersive coal dust particles of the small sizes in our experiments, hence there were no the conditions of the dense structural barrier creation, which is usually made of the small dispersive coal dust particles of the large sizes in the sub-surface layer of the granular filtering medium of the absorber, blocking the air-dust aerosol stream flow in the vertical *IAF*. In our experiments, the appearing weakly penetrating dense barrier, which was created by the small dispersive coal dust particles of the micro- and nano- sizes in the sub-surface layer of the granular filtering medium in an absorber in a model of the vertical *IAF*, was not dense enough to completely block the air-dust aerosol stream flow in the vertical *IAF*. This is a main difference, comparing to a case of the structural barrier creation by the small dispersive coal dust particles of the relatively large sizes in the sub-surface layer of the granular filtering medium in the absorber, which was dense enough to almost completely block the air-dust aerosol stream flow in the real *IAF* at the *NPP*. In our experiment, the air-dust aerosol stream flow had a the high enough velocity in the granular filtering medium inside an absorber, resulting in the transportation of the small dispersive coal dust particles of the very small sizes and masses, and their subsequent jettisoning outside a model of the vertical *IAF*. The maximums of the small dispersive coal dust particles of the very small masses were not detected in the researched case, because of the above reason.

In the course of the conducted experimental research, it is found that the shift of the fractional consistence of the small dispersive coal dust particles to a range of the micro- and nano- sizes has a considerable effect on the physical character of the coal dust masses transportation by the air-dust aerosol in the chemical absorber, changing its aerodynamic parameters and prolonging its operational time period in a model of the vertical *IAF* at the *NPP*.

Authors are very grateful to a group of scientists from the *National Academy of Sciences in Ukraine* (*NASU*) for the numerous scientific discussions on the reported experimental research results.

This innovative research is completed in the frames of the nuclear science and technology fundamental research program, facilitating the environment protection from the radioactive contamination, at the *National Scientific Centre Kharkov Institute of Physics and Technology* (*NSC KIPT*) in Kharkov in Ukraine. The research is funded by the *National Academy of Sciences in Ukraine* (*NASU*).

This research paper was published in the *Problems of Atomic Science and Technology* (*VANT*) in 2009 in [4].

*E-mail:  ledenyov@kipt.kharkov.ua .